\documentclass[aps,prb,twocolumn,groupedaddress,showpacs,floatfix,superscriptaddress,longbibliography, nofootinbib]{revtex4-2}

\usepackage[colorlinks = true,
            linkcolor = red,
            urlcolor  = red,
            citecolor = blue,
            anchorcolor = blue]{hyperref}
\usepackage{epsfig}
\usepackage{amsmath,amssymb}
\usepackage{physics}
\usepackage{graphicx}
\usepackage[dvipsnames,usenames]{xcolor}
\usepackage[normalem]{ulem}
\usepackage{todonotes}
\usepackage{soul} 
\usepackage{orcidlink}
\usepackage{glossaries}
\newacronym{CDW}{CDW}{charge-density-wave}
\newacronym{eph}{$e$-ph}{electron-phonon}
\newacronym{DQMC}{DQMC}{determinant quantum Monte Carlo}
\newacronym{QMC}{QMC}{quantum Monte Carlo}
\newacronym{SSH}{SSH}{Su-Schrieffer-Heeger}
\newacronym{HMC}{HMC}{hybrid Monte Carlo}
\newacronym{AFM}{AFM}{antiferromagnetism}
\newacronym{BOW}{BOW}{bond-order-wave}
\newacronym{VBS}{VBS}{valence bond solid}
\newacronym{QCP}{QCP}{quantum critical point}
\newacronym{SOM}{SOM}{supplementary online materials}

\begin{document}

\title{The optical {S}u-{S}chrieffer-{H}eeger model on a triangular lattice}
\author{Max Casebolt}
\affiliation{Department of Physics and Astronomy, University of California, Davis, California 95616, USA}
\author{Sohan~{Malkaruge~Costa}\orcidlink{0000-0002-9829-9017}}
\affiliation{Department of Physics and Astronomy, The University of Tennessee, Knoxville, Tennessee 37996, USA}
\affiliation{Institute for Advanced Materials and Manufacturing, The University of Tennessee, Knoxville, Tennessee 37996, USA\looseness=-1}
\affiliation{Department of Physics, University of Colombo, Colombo 00300, Sri Lanka}
\author{\mbox{Benjamin Cohen-Stead}\orcidlink{0000-0002-7915-6280}}
\affiliation{Department of Physics and Astronomy, The University of Tennessee, Knoxville, Tennessee 37996, USA}
\affiliation{Institute for Advanced Materials and Manufacturing, The University of Tennessee, Knoxville, Tennessee 37996, USA\looseness=-1}
\author{Richard Scalettar\orcidlink{0000-0002-0521-3692}}
\affiliation{Department of Physics and Astronomy, University of California, Davis, California 95616, USA}
\author{Steven Johnston\orcidlink{0000-0002-2343-0113}}
\affiliation{Department of Physics and Astronomy, The University of Tennessee, Knoxville, Tennessee 37996, USA}
\affiliation{Institute for Advanced Materials and Manufacturing, The University of Tennessee, Knoxville, Tennessee 37996, USA\looseness=-1}

\begin{abstract}
We study the triangular lattice optical \gls*{SSH} model using determinant quantum Monte Carlo. By varying the model's carrier concentration $\langle n \rangle$, electron-phonon coupling strength, and phonon energy $\Omega$, we identify two doping regimes of interest. At one-quarter filling ($\langle n\rangle = 0.5$), corresponding to the case of a circular noninteracting Fermi surface, we find evidence for a metal to insulating \gls*{BOW} phase transition that breaks a local $C_6$ rotational symmetry. Conversely, at three-quarters filling ($\langle n\rangle = 1.5$), corresponding to a hexagonal Fermi surface, we find evidence for transitions to a metallic \gls*{BOW} phase for small $\Omega$ and an $s$-wave superconducting phase for sufficiently large $\Omega$. This tendency toward pairing appears to be associated with the possibility of a sign change in the effective nearest-neighbor intersite hopping, which can occur for sufficiently large lattice displacements. We also find no evidence for enhanced magnetic correlations in the model, contrary to what has been reported for some square lattice \gls*{SSH} models. 
\end{abstract} 

\glsresetall
\maketitle

\section{Introduction}
Unraveling the interplay between electrons and phonons is crucial to understanding many-body systems. By allowing electrons to affect the physical space through which they propagate via distortions of the lattice, one can stabilize  states of matter, including phonon-mediated  superconductivity~\cite{BCS, Bradley2021superconductivity, cai2023hightemperature}, bond/charge order waves~\cite{Gruner1988cdw, Costa2020phase, gotz2022valence, MalkarugeCosta2024Kukule}, polaronic states~\cite{Devreese2009Froelich, Franchini2021polarons, Jiang2021polaron, Naamneh2025persistence}, and beyond. 

The \gls*{SSH} model provides a simple yet effective framework for examining how the inclusion of phonons can affect electronic behavior. First put forth by Bari\v si\'c, Labb\'e, and Friedel to study superconductivity in three-dimensional transition metals~\cite{Barisic1970tightbinding}, it is most well known for modeling dimerization in polyacetylene~\cite{Su1979solitons}. The \gls*{SSH} model builds an effective framework for \gls*{eph} interactions by allowing the nearest-neighbor hopping integrals to explicitly depend on the atomic positions, which is then treated by expanding the hopping to linear order in the atomic 
displacement $\boldsymbol{R}_{\boldsymbol{i}} = \boldsymbol{R}^0_{\boldsymbol{i}}+\boldsymbol{Q}_{\boldsymbol{i}}$ such that 
\begin{equation}\label{eq:linear}
t(\boldsymbol{R}_{\boldsymbol{i}} -\boldsymbol{R}_{\boldsymbol{j}}) 
\approx t(\boldsymbol{R}^0_{\boldsymbol{i}} -\boldsymbol{R}^0_{\boldsymbol{j}}) + \boldsymbol{\nabla} t \cdot \left[\boldsymbol{Q}_{\boldsymbol{i}}-\boldsymbol{Q}_{\boldsymbol{j}}\right],  
\end{equation}
where $\boldsymbol{Q}_{\boldsymbol{i}}$ is the atom's displacement from its equilibrium position $\boldsymbol{R}^0_{\boldsymbol{i}}$. The \gls*{SSH} model is thus physically distinct from the widely studied Holstein~\cite{holstein1959} and Fr{\"o}hlich~\cite{frohlich54electrons} models, where the atomic motion modulates the electron site energy. Recent nonperturbative studies of \gls*{SSH}-like models have also found that this coupling mechanism can give rise to richer phenomena that are not typically associated with \gls*{eph} interactions, including phonon-mediated 
\gls*{AFM}~\cite{afmssh}, quantum spin liquids~\cite{cai2024quantum}, nontrivial topological states~\cite{Moeller2017typeII, Li2023topological, DiSalvo2024topological, sousa2026real}, and high-temperature superconductivity~\cite{Zhang2023bipolaronic, TanjaroonLy2023comparative, Cai2025hightemperature}. 

\gls*{SSH} models on the triangular lattice are particularly alluring in this context. This lattice is the simplest structure to realize kinetic energy and geometric frustration, which can now be modulated by the \gls*{eph} interaction. Frustration introduces competition between standard forms of order where one might otherwise dominate the parameter space. For instance, the Holstein model on a half-filled square lattice easily develops charge order due to its bipartite structure and nested Fermi surface and, as such, Cooper pairing is suppressed by bipolaron formation~\cite{Esterlis2018breakdown, Bradley2021superconductivity, Nosarzewski2021superconductivity}. However, the charge order is substantially suppressed when the model is placed on a triangular lattice, leading to $s$-wave superconductivity at smaller \gls*{eph} coupling strengths~\cite{frustration}. Similarly, \textit{bond} \gls*{SSH} models on half-filled square lattices have \gls*{BOW} or \gls*{AFM} ground states~\cite{afmssh, gotz2022valence, Feng2022phase} while the same interaction on a half-filled triangular lattice may stabilize a quantum spin liquid phase~\cite{cai2024quantum}. 

To date, three different variants of single-band \gls*{SSH} models have been studied in the literature, which differ in how they treat the \gls*{eph} coupling and non-interacting phonon terms. They are the bond~\cite{Sengupta2003Peierls}, optical \cite{Capone1997small}, and acoustic~\cite{Barisic1970tightbinding} models, following the nomenclature introduced in Ref.~\cite{MalkarugeCosta2023comparative}. Crucially, these models cannot be straightforwardly mapped onto one another~\cite{MalkarugeCosta2023comparative, TanjaroonLy2023comparative, silva2025twodimensional}. Motivated by this, we present here a \gls*{DQMC} study of the triangular lattice optical \gls*{SSH} model, in contrast to the bond model studied in Ref.~\cite{cai2024quantum}. By systematically varying the model parameters, we identify two filling fractions of interest. At one-quarter filling ($\langle n \rangle = 0.5$), when the noninteracting band has a circular Fermi surface, we find the presence of a robust insulating \gls*{BOW} phase 
in the weak coupling regime. Conversely, for three-quarter filling ($\langle n \rangle = 1.5$), where the noninteracting band has a hexagonal Fermi surface, we find evidence for $s$-wave superconductivity in the antiadiabatic limit ($\Omega/t > 1.5$) and a metallic \gls*{BOW} in the adiabatic limit. These phases appear to be in competition with one another and separated by a metallic region in parameter regimes where the linear approximation [Eq.~\eqref{eq:linear}] is valid. \\

\section{Model \& Methods}
We investigate the optical-\gls*{SSH} model on a triangular lattice described by the Hamiltonian
\begin{equation}
    \begin{aligned}
        \hat{H} = & -\sum_{\boldsymbol{j},\nu,\sigma} \left(t - \alpha (\hat{\boldsymbol{Q}}_{\boldsymbol{j}+\boldsymbol{a}_\nu} - \hat{\boldsymbol{Q}}_{\boldsymbol{j}})\cdot\boldsymbol{a}_\nu \right)\hat{c}^{\dagger}_{\boldsymbol{j}+\boldsymbol{a}_\nu,\sigma} \hat{c}^{\phantom\dagger}_{\boldsymbol{j},\sigma}\\
        & - \mu\sum_{{\boldsymbol{j}},\sigma}\hat{n}_{{\boldsymbol{j}},\sigma} + 
        \sum_{\boldsymbol{j}}\left( 
        \frac{|\hat{\boldsymbol{P}}_{\boldsymbol{j}}|^2}{2M} + 
        \frac{M\Omega^2 |\hat{\boldsymbol{Q}}_{\boldsymbol{j}}|^2}{2}
        \right). 
    \end{aligned}
    \label{eq:H}
\end{equation}
Here, the sum over $\boldsymbol{j}$ runs over all sites in the lattice. The sum over $\nu \in \{ 1, 2, \dots, 6 \}$ runs over six vectors $|\boldsymbol{a}_\nu| = a$ connecting a site to its nearest-neighbors, with $a$ the lattice constant. The operators $\hat{c}^\dagger_{{\boldsymbol{j}},\sigma}/\hat{c}^{\phantom\dagger}_{{\boldsymbol{j}},\sigma}$ are the fermion creation/annihilation operators for a spin $\sigma$ on site ${\boldsymbol{j}}$, $t$ is the nearest neighbor hopping integral, and $\mu$ is the chemical potential.  The motion of the lattice is described by Einstein oscillators in the two spatial directions, with the phonon displacement $\hat{\boldsymbol{Q}}_{\boldsymbol{j}}=\big( \hat{X}_{\boldsymbol{j}}, \hat{Y}_{\boldsymbol{j}} \big)$ and corresponding momentum $\hat{\boldsymbol{P}}_{\boldsymbol{j}}$ operators. Here, $M$ is the ion mass and $\Omega$ is the frequency of the optical phonons. Finally, $\alpha$ sets the strength of the \gls*{eph} coupling, which arises from the (linear) change of the hopping integrals projected onto the nearest-neighbor bond directions. From this point onward, we work in units such that $\hbar = t = M = a = k_\mathrm{B} = 1$ and report our results in terms of a dimensionless \gls*{eph} coupling constant $\lambda = \alpha^2/(M\Omega^2t)$, phonon energy $\Omega$, and average electron filling $\langle n\rangle = \frac{1}{N}\sum_{\boldsymbol{j},\sigma}\langle \hat{n}_{\boldsymbol{j},\sigma}\rangle$. 

\begin{figure}[t]
\centering{
\includegraphics[width=1.1\columnwidth]{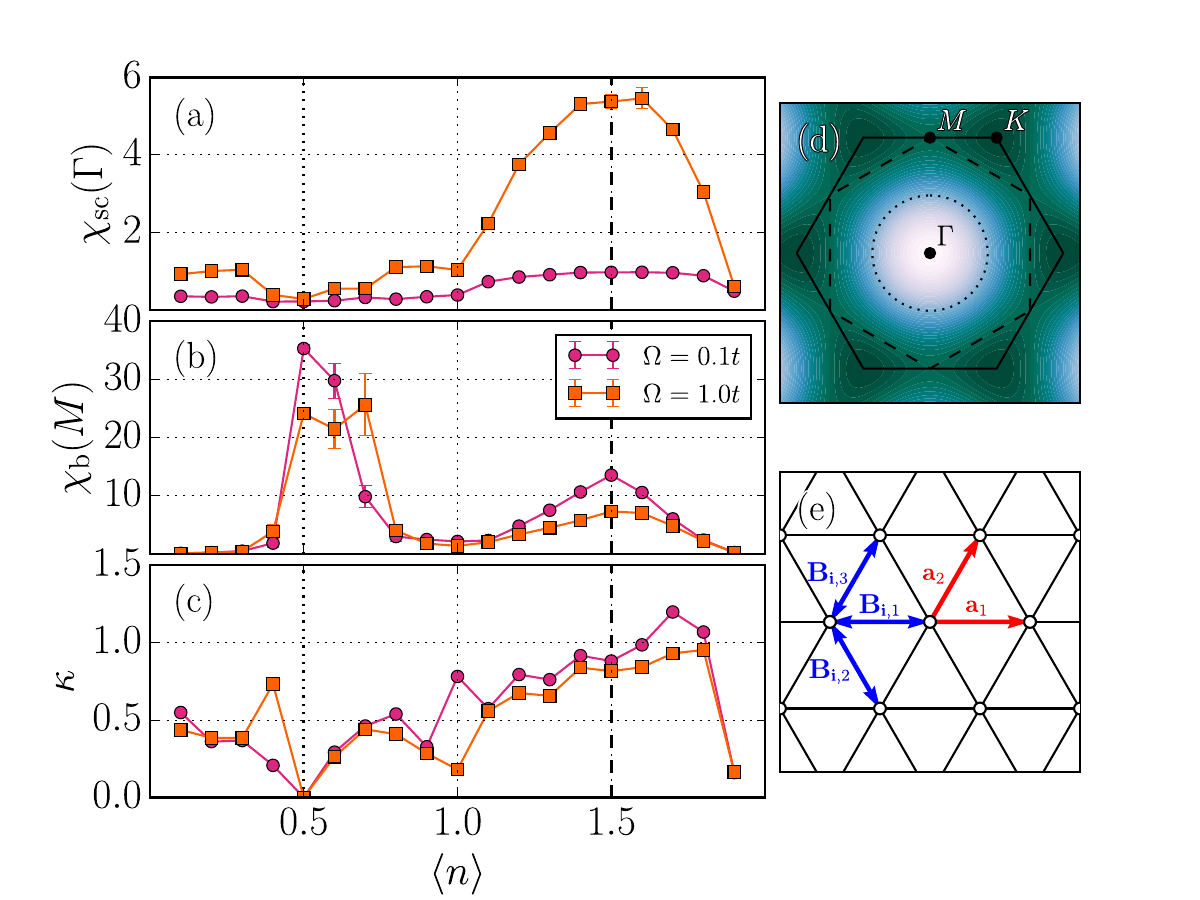}
}
\vspace{-0.75cm}
\caption{Low temperature ($\beta t= 20$) results for the superconducting ($\chi_\text{sc}$, panel a) and bond ($\chi_\text{b}$, panel b) susceptibility, and electronic compressibility ($\kappa$, panel c) as functions of average density $\langle n \rangle$ and fixed $\lambda=\tfrac{\alpha^2}{\Omega^2}=0.25$. Results are shown for $\Omega = 0.1t$ and $1.0t$ as indicated by the common legend. Panel (d) shows the Fermi surfaces of the non-interacting system at $\langle n\rangle = 0.5$ (dotted circle) and $\langle n\rangle=1.5$ (dashed hexagon).  The $\Gamma$, $K$, and $M$
points are also indicated. Panel (e) shows the lattice vectors and the orientation of each of the bond operators used to define the bond correlation functions (see main text).}
\label{fig:doping_sweep} 
\end{figure}

We simulate Eq.~\eqref{eq:H} on $N = L\times L$ lattices with periodic boundary conditions, where $L$ is the extent of the cluster in the direction of each lattice vector (see Fig.~\ref{fig:doping_sweep}).  We use \gls*{DQMC} simulations with \gls*{HMC} updates~~\cite{Beyl2018revisiting, Batrouni2019langevin, CohenStead2022fast}, as implemented in the \texttt{SmoQyDQMC.jl} package~\cite{SmoQyDQMC1, SmoQyDQMC2}.  The \gls*{DQMC} methodology is free of the sign-problem since, for the \gls*{SSH} Hamiltonian, the two spin species couple symmetrically to the phonon coordinates, so their (real-valued) determinants are equal and the weight is a perfect square. 

All simulations were performed with 12-40 parallel Markov chains and  $5\times10^3$ thermalization updates. The number of measurement updates ranged between $1\times10^3$ to $1\times10^4$, depending on the number of parallel walkers used. To maintain comparable statistical accuracy over all simulations, we increased the number of parallel walkers when fewer measurement updates were performed. All measurements were binned with $100-500$ measurement updates averaged per bin. In all simulations, we performed \gls*{HMC} updates with $N_t=8$ time steps of size $\Delta t=\pi/(2\Omega N_t)$ with a small amount of jitter. We refer the reader to Ref.~\cite{SmoQyDQMC1} for additional details. 

To assess the strength of the superconducting correlations, we measured the $s$-wave pair-field susceptibility
\begin{equation}
    \chi_\text{sc}(\boldsymbol{q}) = \frac{1}{N}\sum_{\boldsymbol{i}, \boldsymbol{j}}\int_0^\beta d\tau 
    \langle \hat \Delta_{\boldsymbol{i}}^{\phantom{\dagger}}(\tau)
    \hat \Delta_{\boldsymbol{j}}^{\dagger}(0)
    \rangle e^{-\mathrm{i}\boldsymbol{q}\cdot(\boldsymbol{R}^0_{\boldsymbol{i}}-\boldsymbol{R}^0_{\boldsymbol{j}})},
    \label{eq:chisc}
\end{equation}
where $\hat \Delta_{\boldsymbol{i}}(\tau)=e^{\tau \hat{H}} \hat{c}_{\boldsymbol{i},\uparrow} \hat{c}_{\boldsymbol{i},\downarrow} e^{-\tau \hat{H}}$ is the imaginary time-dependent pairing operator. 

Previous studies of \gls*{SSH} models have observed \gls*{BOW} phases that break various rotational symmetries~\cite{afmssh, cai2024quantum, TanjaroonLy2023comparative, MalkarugeCosta2024Kukule, MalkarugeCosta2023comparative, gotz2022valence}. Therefore, we expect formation of \gls*{BOW} phases that break either $C_2$, $C_3$, or $C_6$ rotational symmetries on a triangular lattice. To observe such transitions we introduce order parameters
\begin{equation}
    \hat{\Psi}_n(\boldsymbol{q})=\frac{1}{N}\sum_{\boldsymbol{j},\nu} \left[e^{-\mathrm{i}\boldsymbol{q}\cdot \boldsymbol{R}^0_{\boldsymbol{j}}}e^{\mathrm{i}2\pi\nu /n}\hat{B}_{\boldsymbol{j},\nu}\right],
\end{equation}
with $n\in \{ 2, 3, 6 \}$ signifying a \gls*{BOW} with a broken $C_n$ rotation symmetry and ordering wavevector $\boldsymbol{q}$. The bond operator is represented by 
$\hat B_{\boldsymbol{j},\nu}  = \sum_{\sigma}\left[\hat c^\dagger_{\boldsymbol{j},\sigma}\hat c^{\phantom\dagger}_{\boldsymbol{j}+\boldsymbol{a}_\nu,\sigma}+h.c.\right]$,  
where $\boldsymbol{a}_\nu$ is again the displacement to the nearest neighbor site [see Fig.~\ref{fig:doping_sweep}(e)].  The equal time \gls*{BOW} structure factor is then defined as $S_n^\text{vbs}(\boldsymbol{q})=N|\hat{\Psi}_n(\boldsymbol{q})|^2$. 
We will also compute the associated susceptibility $\chi_{n}({\boldsymbol q})$ obtained by using imaginary time displaced bond operators 
$\hat B_{\boldsymbol{j},\nu}(\tau)=e^{\tau \hat{H}} \hat B_{\boldsymbol{j},\nu} e^{-\tau \hat{H}}$ 
and integrating over all imaginary time displacements, in analogy with the pairing susceptibility of Eq.~\eqref{eq:chisc}.

To assess the strength of spin correlations, we measured the spin-spin correlation function 
\begin{equation}
    S(\boldsymbol r) = \frac{1}{N}\sum_{\boldsymbol{j}} \langle \hat S^z_{\boldsymbol{j}}\hat S^z_{\boldsymbol{j} + \boldsymbol{r}}\rangle,
\end{equation}
where $\hat S^z_{\boldsymbol{j}} =  \left[\hat{n}_{\boldsymbol{j,\uparrow}}-\hat{n}_{\boldsymbol{j,\downarrow}}\right]$ is the local spin-$z$ operator. The corresponding structure factor is defined as
\begin{equation}
    S(\boldsymbol{q}) = \sum_{\boldsymbol r} e^{-\mathrm{i}\boldsymbol q\cdot \boldsymbol{r}}S(\boldsymbol{r}).
\end{equation}

To estimate the superconducting or \gls*{BOW} critical temperatures $T_\mathrm{c}$, we apply the correlation ratio method~\cite{kekule-op2}. Specifically, we compute 
\begin{equation}
    \mathcal R_\Theta(\boldsymbol{q}) = 1 - \frac{\sum_m{\Lambda_\Theta(\boldsymbol{q}+\delta \boldsymbol{q}_m)}}{6\Lambda_\Theta(\boldsymbol{q})}, 
\end{equation}
where $\Theta=\text{bow}/\text{sc}$ labels the phase and $\Lambda$ is the corresponding structure factor or susceptibility. Since we expect a uniform superconducting state, we chose the pair-field susceptibility $\Lambda\equiv\chi_\text{sc}(\boldsymbol{q}=\boldsymbol{\Gamma})$ for $\mathcal R_\text{sc}$. Conversely, 
for the \gls*{BOW} correlation ratio, we chose the equal time \gls*{BOW} structure factor $\Lambda \equiv S_{n}^\text{bow}(\boldsymbol{q}=\boldsymbol{M})$. Note, we have not observed any strong signals for any other momentum points, indicating that the dominate ordering appears at these wave vectors. In both cases, $\delta \boldsymbol{q}_m$ are the vectors between $\boldsymbol{q}$ and its nearest-neighbors points in momentum space. 
In the ordered phase, we expect a sharp peak to form at ordering vector $\boldsymbol{q}$, creating a 
large difference between $\Lambda_\Theta(\boldsymbol{q})$ and $\Lambda_\Theta(\boldsymbol{q}+\delta\boldsymbol{q}_m)$. The correlation ratio will approach 1 in this case. Conversely, in the unordered phase, the structure factor will be broad and the ratio is approximately zero. Importantly, $\mathcal R_\Theta(\boldsymbol{q})$ is renormalization invariant and will cross at a critical point for different system sizes \cite{Binder1981finite, Kaul2015spin, Pujari2016interaction, kekule-op2}, thereby providing an estimate for the location of critical points in the thermodynamic limit. 

Finally, we estimated for the total density of states at the Fermi energy $N(0)$ in some cases by measuring the local Green's function at $\tau = \beta/2$ 
\begin{equation}\label{eq:N0}
    N(0) \approx \frac{\beta}{\pi}G(\boldsymbol{r} = 0, \tau = \beta/2). 
\end{equation}
See Ref.~\cite{Trivedi1995deviations} for details.

\section{Results}
We begin by examining the superconducting and \gls*{BOW} correlations as a function of doping and phonon energy. 
Figure~\ref{fig:doping_sweep} plots the evolution of the ($s$-wave) superconducting and bond susceptibilities at low temperature ($\beta t = 20$). For small phonon energies in the adiabatic limit ($\Omega = 0.1t$), the pairing correlations at all carrier concentrations remain small, but are slightly enhanced near three quarter filling ($\langle n \rangle = 1.5$). Note that this filling value corresponds to a hexagonal Fermi surface in the noninteracting limit [see Fig.~\ref{fig:doping_sweep}(d)]. The pairing correlations are enhanced significantly when the phonon energy is increased to $\Omega=t$, consistent with expectations for a phonon-mediated pairing mechanism. 

\begin{figure}[t]
\centering{
\includegraphics[width=\columnwidth]{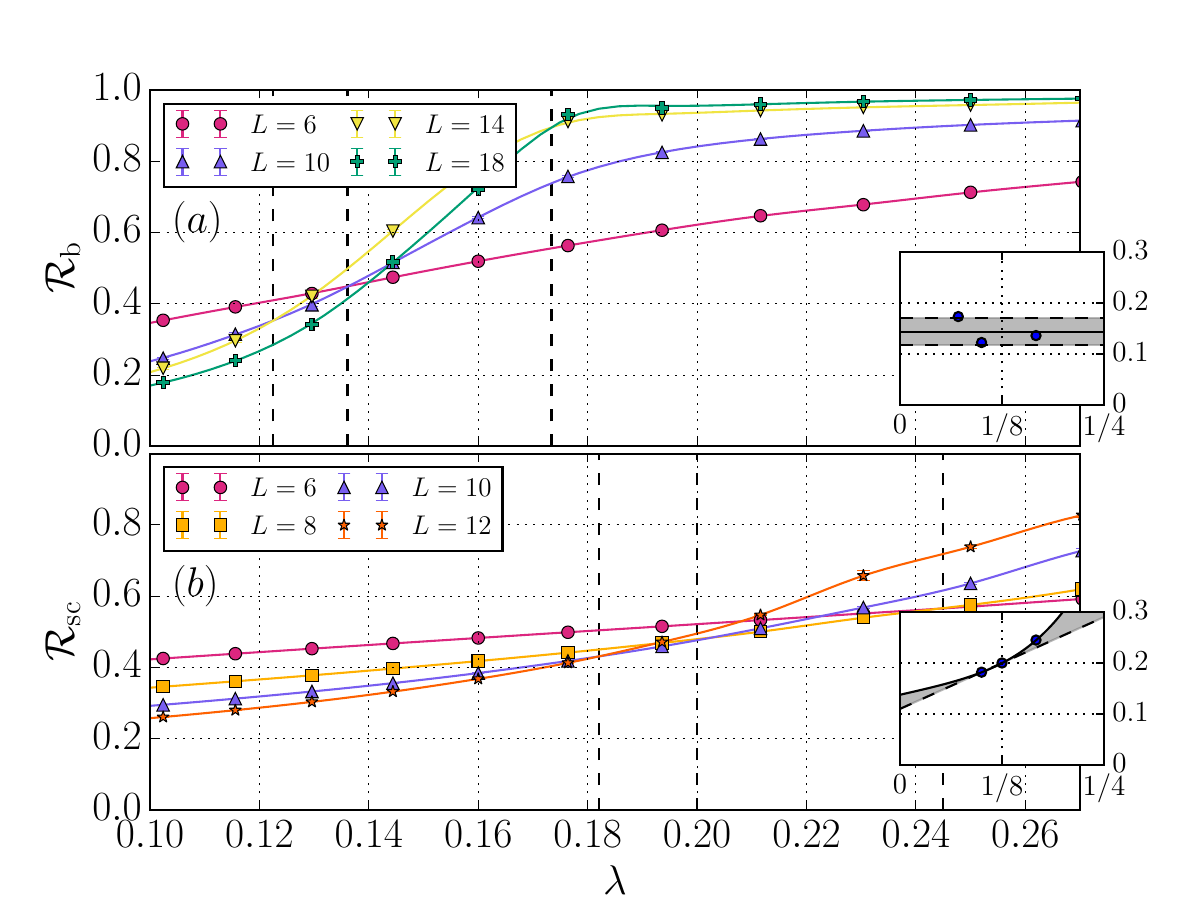}
}
\vspace{-0.25cm}
\caption{The $\lambda$-dependent finite size behavior of the bond-wave $\mathcal{R}_\text{b}$ ($\langle n\rangle=0.5$, $\Omega=0.1t$) and superconducting $\mathcal{R}_\text{sc}$ ($\langle n\rangle=1.5$, $\Omega=2t$) ratios when setting $L=\beta t$.  Different methods were used to estimate the critical values in the thermodynamic limit 
from plotting the crossings as a function of $1/L$, as shown in the inset figures.
}
\label{fig:Rcollapse} 
\end{figure}

\begin{figure}
    \centering
    \includegraphics[width=\linewidth]{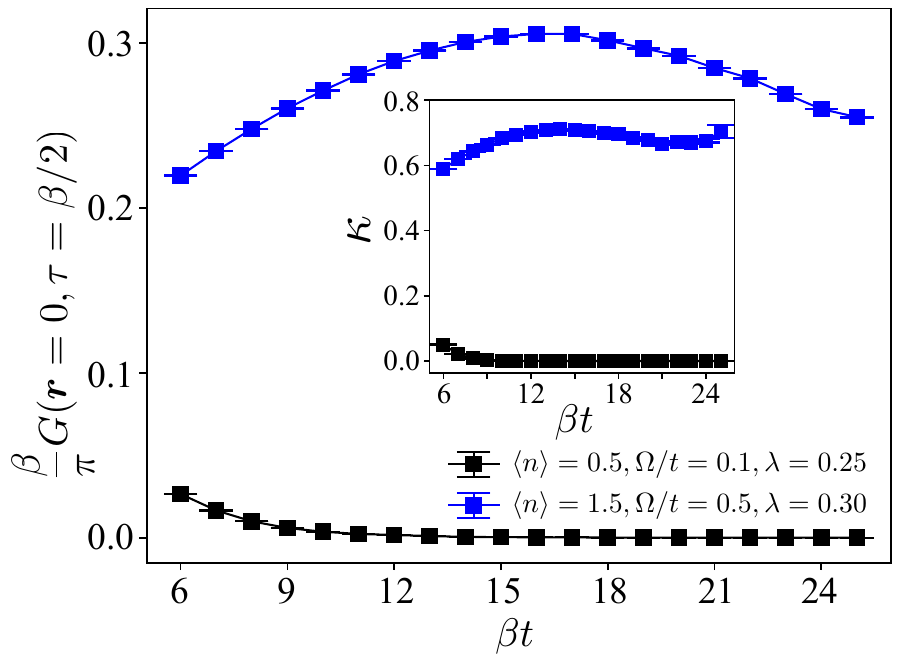}
    \caption{
    Temperature dependence of the spectral weight at the Fermi surface, 
    estimated from the local Green's function [see Eq.~\ref{eq:N0}], as a function of temperature. Results are shown in the BOW phases found at $\langle n \rangle = 0.5$ (black) and 1.5 (blue). Inset: Evolution of the electronic compressibility, $\kappa$, as a function of inverse temperature. All results were obtained on $L = 8$ clusters.
    }
    \label{fig:N0_BOW}
\end{figure}

Figure~\ref{fig:doping_sweep}(b) plots the corresponding evolution of the dominant bond correlations, which appear at the $M$-point. We find that they are strongest for fillings near $\langle n\rangle = 0.5$ and $1.5$, with the dominant correlations appearing at one quarter filling. These filling values correspond to a circular
and hexagonal noninteracting Fermi surface, respectively,
as sketched in Fig.~\ref{fig:doping_sweep}(d).  The correlations at $\langle n\rangle = 0.5$ are large for both values of $\Omega$, and correspond to a broken $C_6$ rotational symmetry of the lattice. We also find that the electronic compressibility $\kappa = \frac{1}{ \langle n \rangle^2}\frac{\partial \langle n \rangle}{\partial \mu}\big|_T$, shown in Fig.~\ref{fig:doping_sweep}(c), goes to zero at this filling, indicating that the \gls*{BOW} state is insulating (see also Fig.~\ref{fig:N0_BOW}). The bond correlations at $\langle n\rangle = 1.5$ are comparatively weaker, and diminish in strength as the phonon energy increases. Notably, the bond correlations are clearly competing with the superconducting correlations in the adiabatic limit and at the larger filling; $\chi_\mathrm{sc}(\Gamma) \approx \chi_\mathrm{b}(\boldsymbol{M})$ when $\Omega = t$ but $\chi_\mathrm{sc}(\Gamma) < \chi_\mathrm{b}(\boldsymbol{M})$ when $\Omega$ is reduced to $0.1t$. The electronic compressibility at this filling is also nonzero, suggesting that the \gls*{BOW} phase is metallic.

The results shown in Fig.~\ref{fig:doping_sweep} suggest that $\langle n \rangle = 0.5$ and $1.5$ are special filling fractions for the optical \gls*{SSH} model on the triangular lattice. We, therefore, focus on these two values for the remainder of this discussion. For both filling values, we estimate the locations of the zero-temperature phase boundaries by carrying out a finite size scaling analysis of the relevant correlation ratios, as exemplified in Fig.~\ref{fig:Rcollapse}. In all cases, we scale the inverse temperature with the lattice size to maintain $L = \beta t$ to capture the growth of temporal fluctuations at low temperatures. This choice assumes that the correlation ratio is Lorentz invariant at the \gls*{QCP}, with a dynamical critical exponent of $z = 1$. In doing so, we can estimate the critical couplings $\lambda_\mathrm{c}$ for the \gls*{BOW} and superconducting quantum phase transitions. 

Figure~\ref{fig:Rcollapse}(a) demonstrates the correlation ratio analysis for the \gls*{BOW} phase transition. Here, we estimate the critical coupling $\lambda_\mathrm{c}(L)$ for a given cluster size $L$ from the crossing point in the correlation ratio curve obtained using clusters of size $L$ and $L+2$. Plotting the results as a function of $1/L$, as shown in the inset, shows that the crossing values fluctuate around $\lambda_\mathrm{c} \approx 0.129\pm0.01$. We take this value as an estimate of the critical coupling in the thermodynamic limit while using the spread of the values as a measure of the uncertainty. Fig.~\ref{fig:Rcollapse}(b) demonstrates a similar analysis for the superconducting phase transition. Unlike in Fig.~\ref{fig:Rcollapse}(a), here, we observe a much smoother trend in the crossing points as the cluster size is increased (see also the inset). In this case, we extrapolate the line going through these points to $1/L = 0$ and obtain $\lambda_\mathrm{c} = 0.128\pm0.012$ as an estimate for the superconducting transition in the thermodynamic limit.

\begin{figure}[t]
\centering{
\includegraphics[width=\columnwidth]{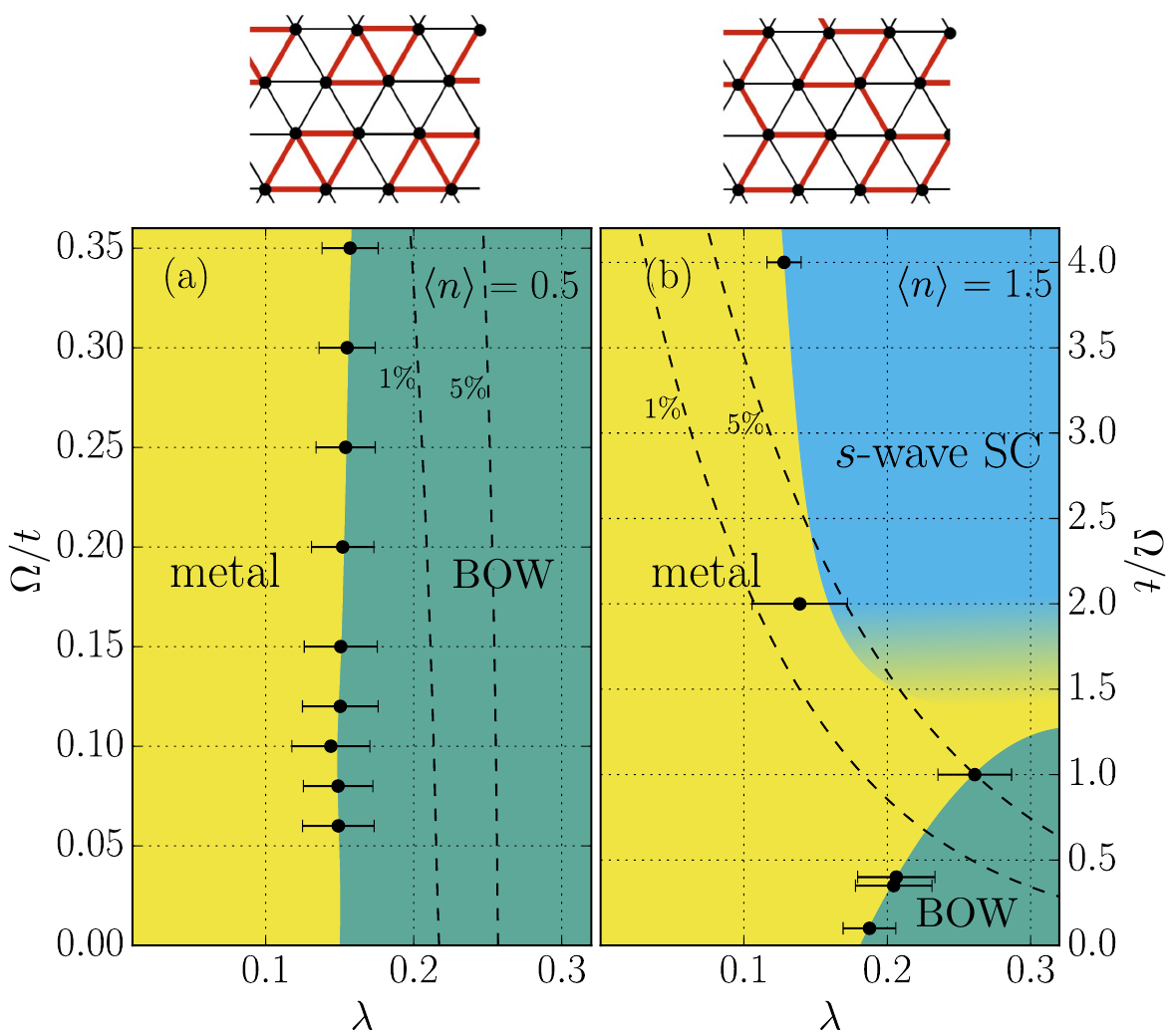}
}
\vspace{-0.5cm}
\caption{The ground state $\lambda-\Omega$ phase diagrams of the triangular lattice optical \gls*{SSH} model at (a) $\langle n\rangle=0.5$  and (b) $\langle n\rangle=1.5$. At the lower density there is a metal-\gls*{BOW} insulator transition with increasing \gls*{eph} coupling. At higher density superconductivity prevails over \gls*{BOW} order for sufficiently large $\Omega$. 
The shading in the lower boundary between the superconducting and metallic phases is meant to highlight  uncertainty in the presence and location of the phase boundary given our current numerical results. The left and right dashed contours correspond to cases where sign flipping appears in the 1\% and 5\% of the measurements, respectively. The lattice diagrams shown above each panel indicate the displacement patterns associated with each of the \gls*{BOW} phases.}
\label{fig:phase_diagram} 
\end{figure}

Based on similar data and finite-size scaling analyses of the correlation ratios, we determined the model's low temperature phase diagrams in the $\lambda-\Omega$ plane, as shown in Fig.~\ref{fig:phase_diagram}. 
At $\langle n \rangle = 0.5$ [Fig.~\ref{fig:phase_diagram}(a)], the system undergoes a metal to \gls*{BOW} phase transition with broken $C_6$ symmetry. The critical coupling depends linearly on the phonon energy $\Omega$ such that larger $\alpha$ values are needed to establish the \gls*{BOW} phase as $\Omega$ increases.  This behavior is reflected in the nearly vertical phase boundary in Fig.~\ref{fig:phase_diagram}(a), where $\lambda$ is used as the horizontal axis. Note, ``metal'' in Fig.~\ref{fig:phase_diagram} refers to a region  where we are unable to resolve any broken symmetries within the accessible temperatures and the limitations of our numerical simulations. Given the absence of a sign problem for this model, one expects that the system's ground state is not a conventional Fermi liquid~\cite{Grossman2023robust} and we cannot completely rule out the possibility a superconductor, magnetic order, or a non-Fermi-liquid-like state at lower temperatures.

The phase diagram at three quarter filling [Fig.~\ref{fig:phase_diagram}(b)] is somewhat richer. In this case, we observe a metal-to-\gls*{BOW} phase transition for $\Omega/t \lessapprox 1.25$ and superconducting transition for sufficiently large values of $\Omega/t$. We are unable to resolve a superconducting or \gls*{BOW} transition for intermediate phonon energies ($\Omega/t \approx 1.5)$ within temperatures and cluster sizes accessed here. For this reason, we could not locate the precise location of the lower boundary of the superconducting phase. The color gradient in Fig.~\ref{fig:phase_diagram} highlights this uncertainty. 

In simulations of the optical \gls*{SSH} model, the lattice displacements can become large enough to flip the sign of the electronic hopping, which reflects a breakdown of the linear approximation used in Eq.~\eqref{eq:linear}~\cite{Nocera2021bipolaron, Banerjee2023ground}. For this reason, we monitor the percentage of times that the sign of one of the effective hopping integrals $t_\mathrm{eff} \approx t - \alpha (\langle\hat{\boldsymbol{Q}}_{\boldsymbol{j}+\boldsymbol{a}_\nu} - \hat{\boldsymbol{Q}}_{\boldsymbol{j}}\rangle)\cdot\boldsymbol{a}_\nu $ inverts in our simulations. The dashed lines in Fig.~\ref{fig:phase_diagram} show contours for 1 and 5\% sign changes. For $\langle n \rangle = 0.5$, this sign switching occurs deeper in the \gls*{BOW} phase. Conversely, for $\langle n\rangle = 1.5$, it appears near the boundary of the superconducting phase and upper portion of the \gls*{BOW} phase. This result suggests that the superconductivity observed at large $\Omega/t$ may be difficult to realize in a real system where anharmonic lattice potentials and nonlinear \gls*{eph} coupling would be naturally present for sufficiently large lattice displacements.

\begin{figure}[t]
\centering{
\includegraphics[width=1.1\columnwidth]{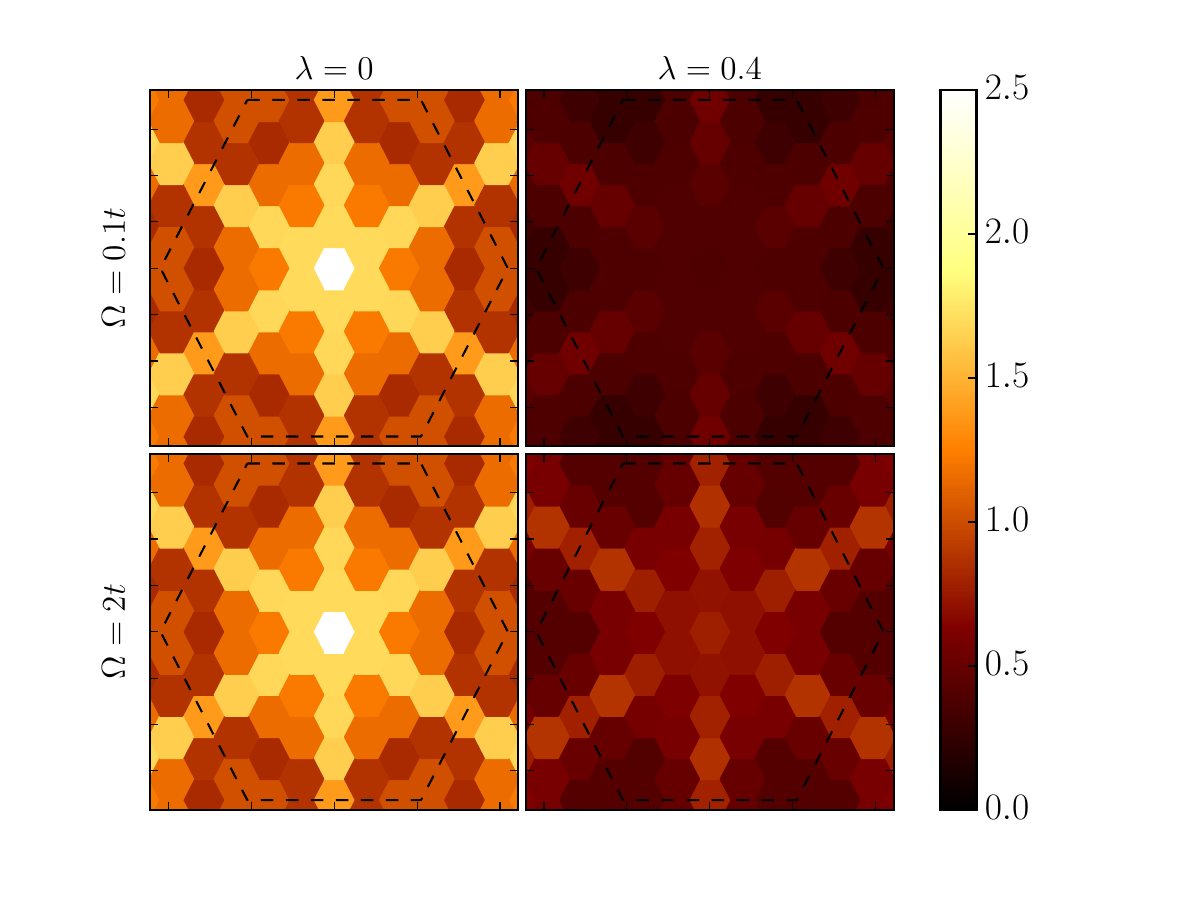}
}
\caption{A heat map of the spin structure factor for $\langle n\rangle=1.5$ $L=8$ at $\Omega=0.1t, 2t$ and $\lambda=0, 0.4$.  Increasing the dimensionless coupling electron-phonon suppresses spin correlations at all momenta
and also in  both the adiabatic and anti-adiabatic regimes.  Thus magnetism appears not to play a role in this
geometry, in contrast to the SSH model on a square lattice.
}
\label{fig:Sspin} 
\end{figure}

Given their proximity to the sign-switching region of the model, both \gls*{BOW} phases are likely in the model’s strong-coupling regime. In that case, the \gls*{BOW} order can result in a fully gapped system, even if the underlying noninteracting Fermi surface is not well nested. Fig.~\ref{fig:doping_sweep}(c) already suggests that the \gls*{BOW} phase at one-quarter filling is insulating while the one at three-quarter filling is metallic. To confirm this behavior, Fig.~\ref{fig:N0_BOW} plots the temperature dependence of the spectral weight at the Fermi surface $N(0)$ [see Eq.~\ref{eq:N0}] and electronic compressibility $\kappa$ at points in parameter space located in each of the \gls*{BOW} phases. At one-quarter filling, we find clear indications of insulating behavior: the spectral weight decreases with increasing $\beta$ and is completely suppressed by $\beta t\gtrsim 10$. At the same time, the electronic compressibility falls to zero in the same temperature region, consistent with a full gap in the single-particle density of states and insulating behavior. Conversely, the behavior at three-quarters filling is consistent with a partial gap or metallic conduction. Both the spectral weight and compressibility remain large and are weakly dependent on temperature, down to the highest values of $\beta$ measured.  

Finally, we turn to the possible question of magnetism in the model. This analysis is motivated by \gls*{QMC} simulations on a square lattice, which have shown that weak \gls*{AFM} order is also present in the bond \gls*{SSH} model~\cite{afmssh,gotz2022valence}. These magnetic correlations can be traced to a positive effective intersite exchange $J$ that appears when the phonons are integrated out of the model. At the same time, the tendency toward \gls*{AFM} in the optical \gls*{SSH} model is comparatively weaker, which has been linked to the presence of longer-range exchange couplings that are mediated by the coordinated modulation of the neighbor hopping integrals~ \cite{ly2025antiferromagnetic}. In light of this, it is natural to ask whether any significant magnetic correlations develop on the triangular lattice geometry considered here. This question is also interesting in the context of a recent projector \gls*{QMC} study that found evidence for a quantum spin liquid state in the half-filled bond \gls*{SSH} model~\cite{cai2024quantum}. To address this question, Fig.~\ref{fig:Sspin} plots the 
magnetic structure factor in the first Brillouin zone for $\langle n \rangle = 1.5$. In this case, results are shown for $\Omega = 0.1t$ (top row) and $2t$ (bottom row) with $\lambda = 0$ (left column) and $0.4$ (right column). At this filling, the noninteracting spin-spin correlations have a peak close to the $M$ points, consistent with nesting across the hexagonal noninteracting Fermi surface. However, introducing a nonzero $\lambda$ dramatically suppresses the spin response, indicating that magnetic correlations in the interacting model are not enhanced. This result highlights a key different between the triangular geometry considered here and the square lattice.  

\section{Discussion}
We have studied the optical variant of the \gls*{SSH} model on a triangular lattice, where the atomic motion modulates the hopping integrals. The microscopic nature of the coupling is such that it directly couples to the degree of kinetic energy frustration generated by the triangular lattice geometry. Our key results 
are the demonstration of a \gls*{BOW} state for both quarter and three-quarter filling in the adiabatic limit, and a superconducting phase at $\langle n \rangle = 1.5$ for sufficiently large phonon energies, exceeding the value of the bare hopping $t$. Notably, the superconducting phase also appears to occur largely in a region of parameter space where the linear approximation for the \gls*{SSH} coupling is beginning to break down. In the future, it would be interesting to explore how nonlinear \gls*{eph} interactions modify these phase boundaries. 

On bipartite lattices the competition of ordered phases is well-studied; 
the suppression of $d$-wave pairing by the development of antiferromagnetism
in the half-filled Hubbard model or the suppression of $s$-wave superconductivity by charge-density-wave correlations in the half-filled Holstein model are prominent examples. Exploration of non-bipartite lattices, such as the triangular geometry
investigated here, adds a further richness to the problem, as frustration emerges as a third player in the determination of the dominant low temperature order. For example, a natural next study would be to systematically study the effects of doping away from the dominant \gls*{BOW} phases obtained here. The broad peaks observed in the superconducting correlations shown in Fig.~\ref{fig:doping_sweep}, particularly around three-quarters filling, suggests that superconductivity could be stabilized in this regime. \\

\section*{Acknowledgments} We thank C.~D.~Batista for useful discussions. This work was supported by the U.S.~Department of Energy, Office of Science, Office of Basic Energy Sciences, under Award Number DE-SC0022311. \\

\section*{Data Availability} The data supporting this study has been deposited in an open online repository~\cite{casebolt_2026_22097800}. 

\bibliography{biblio.bib}

\end{document}